\documentclass[jmp,amsmath,amssymb,preprint]{revtex4-1}

\usepackage{graphicx}
\usepackage{dcolumn}

\usepackage[utf8]{inputenc}
\usepackage[T1]{fontenc}
\usepackage{mathptmx}
\usepackage{etoolbox}

\usepackage{mathrsfs}

\DeclareMathOperator{\dive}{div}

\newcommand{\kt}{\rangle}
\newcommand{\br}{\langle}

\newcommand{\pd}{\partial}
\renewcommand{\d}{{\operatorname{d}}}

\newcommand{\R}{\mathbb{R}}
\newcommand{\C}{\mathbb{C}}

\makeatletter
\def\@email#1#2{
	\endgroup
	\patchcmd{\titleblock@produce}
	{\frontmatter@RRAPformat}
	{\frontmatter@RRAPformat{\produce@RRAP{*#1\href{mailto:#2}{#2}}}\frontmatter@RRAPformat}
	{}{}
}%
\makeatother
\begin{document}
	\preprint{AIP/123-QED}
	
	\title[Cylindrical superintegrability with complex magnetic fields]{Cylindrical first order superintegrability with complex magnetic fields}
	\author{O. Kub\r{u}}
	\email{ Ondrej.Kubu@fjfi.cvut.cz, Libor.Snobl@fjfi.cvut.cz}
		\author{L \v{S}nobl}%
	\affiliation{Czech Technical University in Prague, Faculty of Nuclear Sciences and Physical
	Engineering, Department of Physics, B\v{r}ehov\'{a} 7, 115 19 Prague 1, Czech Republic}
	\date{\today}
	
	\begin{abstract}
		This article is a contribution to the study of superintegrable Hamiltonian systems with magnetic fields on the three-dimensional Euclidean space $\mathbb{E}_3$ in quantum mechanics. In contrast to the growing interest in complex electromagnetic fields in the mathematical community following the experimental confirmation of its physical relevance [X. Peng et al., Phys. Rev. Lett. 114 (2015)], they were so far not addressed in the growing literature on superintegrability. 
		Here we venture into this field by searching for additional first order integrals of motion to the integrable systems of cylindrical type. We find that already known systems can be extended into this realm by admitting complex coupling constants. In addition to them, we find one new system whose integrals of motion also feature complex constants. All these systems are multiseparable. Rigorous mathematical analysis of these systems is challenging due to the non-Hermitian setting and lost gauge invariance. We proceed formally and pose the resolution of these problems as an open challenge.
	\end{abstract}
	
	\maketitle
	
	\section{Introduction}
	\label{sec:intro}
	
	This article is a contribution to the study of integrable and superintegrable Hamiltonian systems with magnetic fields on the three-dimensional (3D) Euclidean space $\mathbb{E}_3$ in quantum mechanics, with focus on complex valued magnetic fields. More specifically, we assume a Hamiltonian of the form (using units where $e=-1,\ m=1$)
	\begin{equation}\label{HamMagn}
		H=\frac{1}{2}\left(\vec{p}^2+A_j(\vec{x})p_j+p_j A_j(\vec{x})+A_j(\vec{x})^2\right)+W(\vec{x}),
	\end{equation}
	with implicit summation over repeated indices $j=1,2,3$ (in the whole paper), $\vec{p}=-i\hbar\vec{\nabla}$ is the momentum operator and $\vec{A}=(A_1(\vec{x}),A_2(\vec{x}),A_3(\vec{x}))$ and $W(\vec{x})$ are the vector and electrostatic potentials of the electromagnetic field.
	
	Integrability then entails the existence of two algebraically independent integrals of motion $X_1,X_2$ (further specified below) mutually in involution, i.e.
	\begin{equation}\label{IC}
		[H,X_1]=[H,X_2]=[X_1,X_2]=0.
	\end{equation}
	Superintegrability assumes additional one or two integrals, algebraically independent of each other and the integrals needed for integrability.
	They are usually considered to be polynomials in the momenta $p_j$, for computational feasibility usually of a low order (typically~2).
	
	Integrable (and especially superintegrable) systems are rare and distinguished by the possibility to obtain the solution to their equations of motion in a closed form. They are subsequently invaluable for gaining physical intuition and serve as a starting point for modeling more complicated systems. Finding and classifying these systems is therefore of utmost importance.
	
	{Despite its physical relevance, (super)integrability with magnetic fields was mostly ignored due to computational difficulty. The first systematic result remedying this omission was the article by Shapovalov on separable systems \cite{Shapovalov1972}, followed by the articles in $\mathbb{E}_2$ \cite{Berube,McSween2000}. Subsequent articles in $\mathbb{E}_3$ assumed first order integrals \cite{Marchesiello2015} or separation of variables \cite{Zhalij_2015,Marchesiello2018,BertrandSnobl,Fournier2019}, but some recent articles go beyond separation \cite{Marchesiello2022,Kubu2022}. The non-relativistic quantum case with spin was also investigated, see e.g. the recent article \cite{Yurdusen2021} and references therein.}
	
	In all the cases above the magnetic field was by assumption real. On the other hand, there has recently been a growing interest in imaginary or complex magnetic fields following the experimental confirmation of their physical relevance by observing the Yang-Lee zeros \cite{Peng2015}. A recent paper \cite{Fernandez2022} investigates an exact solution for electron in graphene interacting with a complex magnetic field. A paper by Jaramillo \cite{Jaramillo2015} shows a formal analogy between stable marginally outer trapped surfaces (MOTS) of black holes and non-relativistic charged particle in complex electromagnetic field on closed surfaces. In the mathematical community these systems are considered as pseudo-\cite{Mostafazadeh2002} or quasi-Hermitian \cite{Scholtz1992,dieudonne1961quasi} systems, see also \cite{KS_book,Krejcirik2015,Krejcirik2019} for some more recent work and additional references.
	
	As we have demonstrated in the previous paragraph, complex magnetic fields merit further research. However, the non-Hermitian setting of these systems poses several problems that we do not resolve in this paper and pose them as an open challenge.
	
	The most fundamental one is the rigorous definition of the magnetic Hamiltonians. This is caused by the fact that the time-independent gauge transformation $A'(\vec{x})=A(\vec{x})+\nabla \chi(\vec{x}),$ $W'(\vec{x})=W(\vec{x})$, manifesting itself in the quantum context as the position dependent change of phase of the wave function
	\begin{equation}\label{gauge transf}
		U\psi(\vec{x})=\exp\left(\frac{i}{\hbar}\chi(\vec{x})\right)\psi(\vec{x}),
	\end{equation}
	is unbounded once $\chi(\vec{x})$ has a nonvanishing imaginary part, which may be necessary to fully fix the gauge. In other words, we can fix the real part of the vector potential by a unitary gauge transformation, but the choice of the imaginary part involves an unbounded transformation, which may affect the spectral properties.
	
	Even if this problem is addressed, the spectrum of the corresponding Hamiltonian can be complex, which complicates its physical interpretation as only real spectra are usually measurable. The mathematically correct approach to the problem entails pseudo-\cite{Mostafazadeh2002} or quasi-Hermiticity (or self-adjointness) \cite{Scholtz1992,dieudonne1961quasi}, i.e. the Hamiltonian $H$ satisfies $H^\dagger=\Theta H \Theta^{-1}$ where the metric $\Theta$ and its inverse~$\Theta^{-1}$ are bounded operators, and is self-adjoint with respect to the modified scalar product~$\langle\cdot,\Theta\cdot\rangle$. However, in the more general pseudo-Hermitian case, where the metric $\Theta$ is indefinite, the modified ``scalar product'' is indefinite as well and we deal with the so called Krein space, see e.g. \cite{Bognar2012}. Only when the metric is positive, i.e. in the more strict quasi-Hermitian case, we can talk about a nonstandard representation of usual quantum mechanics. (For bounded operators $H$ these definitions coincide if $0\notin \overline{\{\br \psi,\Theta \psi\kt|\psi\in \mathcal{H},\lVert \psi \rVert=1 \}}$, see \cite{Williams1969} for a proof.) For more detail concerning these notions see \cite{KS_book,Krejcirik2015} and references therein.
	
	Following on our research with real magnetic fields \cite{Kubu2020}, we search for superintegrable system of the cylindrical type with additional first order integrals whose electromagnetic fields are complex functions. In Section \ref{sec cyl} we review cylindrical-type integrable systems, i.e. the corresponding integrals of motion and the magnetic field, in cylindrical coordinates. In Section \ref{sec komplex} we specify the problem of first order superintegrability with complex fields and briefly comment on the calculations leading to our classification. Interested reader can find the details of the computation in the Appendix \ref{app}. The core of our article is the Results section \ref{sec res}. There we analyze the new system with complex integrals of motion in Subsection \ref{sec new sys} and the old systems extended by choosing complex coupling constants in Subsection \ref{sec old systems}. Our concluding remarks are in Section \ref{sec: concl}.
	
	\section{Cylindrical--type system}\label{sec cyl}
	Before we specify the corresponding integrals $X_1, X_2$, we have to introduce the formalism used for magnetic field in curvilinear coordinates in classical mechanics, cf. \cite{Marchesiello2018Sph,Fournier2019}. 
	
	Defining the cylindrical coordinates
	\begin{equation}
		x^1=r \cos(\phi), \quad x^2=r\sin(\phi), \quad x^3=Z,
	\end{equation}
	we represent the vector potential $A$ as a 1-form
	\begin{equation}
		A = A_1 \mathrm{d}x^1 + A_2 \mathrm{d}x^2 + A_3 \mathrm{d}x^3 = A_r \mathrm{d}r + A_\phi \mathrm{d}\phi + A_Z \mathrm{d}Z.
	\end{equation}
	Hence we obtain the following transformations 
	\begin{align}\label{transf p}
		A_1 = \cos(\phi)A_r-\frac{\sin(\phi)}{r}A_\phi, \quad A_2 = \sin(\phi)A_r+\frac{\cos(\phi)}{r}A_\phi, \quad A_3 = A_Z.
	\end{align}
	As components of the canonical 1-form $\lambda=p_j\d x^j$, the momenta $p_j$ transform in the same way and we can define the magnetic. momenta by $p^A_j=p_j+A_j$ in both Cartesian and cylindrical coordinates.
	(We used to call $p^A_j$ the gauge covariant momenta, the terminology coined probably in \cite{Marchesiello2015} but implicitly used as velocities $p_j^A=\dot{x}^j$ since the first papers on superintegrability with magnetic field \cite{Dorizzi_1985}. With complex magnetic field it is no longer appropriate terminology because, as we mentioned in the introduction, the gauge transformation \eqref{gauge transf} may be unbounded.)

	Components of the magnetic field 2-form $B=\mathrm{d}A$ are
	\begin{align}
		\begin{aligned}
			B &= B^1 (\vec{x}) \mathrm{d}x^2 \wedge \mathrm{d}z + B^2 (\vec{x}) \mathrm{d}z \wedge \mathrm{d}x^1 + B^3 (\vec{x}) \mathrm{d}x^1 \wedge \mathrm{d}x^2 \\
			&= B^r (r, \phi, Z) \mathrm{d}\phi \wedge \mathrm{d}Z + B^\phi (r, \phi, Z) \mathrm{d}Z \wedge \mathrm{d}r + B^Z(r, \phi, Z) \mathrm{d}r \wedge \mathrm{d}\phi,
		\end{aligned}
	\end{align}
	which leads to the following transformation
	\begin{align} \label{transformB}
		B^1 (\vec{x}) &= \frac{\cos(\phi)}{r}B^r (r, \phi, Z) - \sin(\phi)B^\phi (r, \phi, Z), \nonumber\\
		B^2 (\vec{x}) &= \frac{\sin(\phi)}{r}B^r (r, \phi, Z) + \cos(\phi)B^\phi (r, \phi, Z), \\
		B^3 (\vec{x}) &= \frac{1}{r}B^Z (r, \phi, Z). \nonumber
	\end{align}
	
	We can use the same formalism and notation in quantum mechanics as well, we just have to quantize the equations and the (properly symmetrized) integrals in Cartesian coordinates and subsequently transform our equations into cylindrical ones. For example, the transformed momenta read
	\begin{equation}
		p_1=-i\hbar \left(\cos(\phi)\pd_{r}-\frac{\sin(\phi)}{r}\pd_{\phi}\right), \ p_2=-i\hbar \left(\sin(\phi)\pd_{r}+\frac{\cos(\phi)}{r}\pd_{\phi}\right), \ p_3=-i\hbar\pd_{Z},
	\end{equation}
	i.e. transformation is the same as \eqref{transf p} upon defining $p_{r,\phi,Z}=-i\hbar \pd_{r,\phi,Z}$.

	We can now introduce integrals of motion of the cylindrical type, i.e. integrals that imply separation of Schr\"odinger (or, classically, Hamilton-Jacobi) equation in the cylindrical coordinates in the limit of vanishing magnetic field $\vec B$. Expressed in the cylindrical coordinates they read
	\begin{equation}
		\begin{split} \label{cyl integrals}
			& X_1=(p_\phi^A)^2+\frac{1}{2}\sum_{\alpha=r,\phi,Z}\left(s_1^\alpha(r, \phi, Z)p_\alpha^A+p_\alpha^A s_1^\alpha(r, \phi, Z)\right)+m_1(r,\phi,Z), \\
			& X_2=(p_Z^A)^2+\frac{1}{2}\sum_{\alpha=r,\phi,Z}\left(s_2^\alpha(r, \phi, Z)p_\alpha^A+p_\alpha^A s_2^\alpha(r, \phi, Z)\right)+m_2(r,\phi,Z).
		\end{split}
	\end{equation}
	The functions $s^{r,\phi,Z}_{1,2},m_{1,2}$ together with the electromagnetic fields $B,W$ are to be determined from the integrability conditions \eqref{IC}.
	
	Since the integrals of motion as well as the Hamiltonian and their commutators are differential operators, the integrability conditions \eqref{IC} can be separated into the coefficients of derivatives which must vanish independently, yielding the so-called determining equations. The second order ones can be solved in terms of 5 auxiliary functions $\rho(r),\sigma(r),\psi(\phi), \tau(\phi),\mu(Z)$ of one variable each that we use in the subsequent calculations, namely
	\begin{align}
		\begin{split}\label{scond}
			s_1^r &=\frac{\mathrm{d}}{\mathrm{d}\phi}\psi(\phi), \quad s_1^\phi=-\frac{\psi(\phi)}{r}-r^2\mu(Z)+\rho(r), \quad s_1^Z=\tau(\phi), \\
			s_2^r&=0, \quad s_2^\phi=\mu(Z), \quad s_2^Z=-\frac{\tau(\phi)}{r^2}+\sigma(r),
		\end{split}\\
		\begin{split} \label{Bcond}
			B^r&=-\frac{r^2}{2}\frac{\mathrm{d}}{\mathrm{d}Z}\mu(Z)+\frac{1}{2r^2}\frac{\mathrm{d}}{\mathrm{d}\phi}\tau(\phi), \quad B^\phi=\frac{\tau(\phi)}{r^3}+\frac{1}{2}\frac{\mathrm{d}}{\mathrm{d}r}\sigma(r), \\
			B^Z&=\frac{-\psi(\phi)}{2r^2}+r\mu(Z)-\frac{1}{2}\frac{\mathrm{d}}{\mathrm{d}r}\rho(r)-\frac{1}{2r^2}\frac{\mathrm{d}^2}{\mathrm{d}\phi^2}\psi(\phi),
		\end{split}
	\end{align}
	cf. \cite{Fournier2019,Kubu2020,Kubu2022a}. We further use primes for derivatives of these functions with respect to their respective independent variable.
	
	\section{First order superintegrability in the cylindrical case with complex fields}\label{sec komplex}
	We are looking for first order superintegrability in the cylindrical case. This means that we impose the integrability conditions \eqref{IC} with the cylindrical integrals \eqref{cyl integrals}. First order superintegrability then amounts to the existence of an additional integral of the form (in Cartesian coordinates)
	\begin{equation}\label{int Y}
		Y= k_1 p_1^A+k_2 p_2^A+k_3 p_3^A+k_4 L_1^A+ k_5 L_2^A+k_6 L_3^A+m(x^1,x^2,x^3).
	\end{equation}
	Here we have the magnetic momenta $p^A_j=p_j+A_j(x^1,x^2,x^3)$ and angular momenta $L^A_j=\epsilon_{jkl} x^k p^A_l.$ (We sum over both indices $k,l$ and $\epsilon_{jkl}$ is the totally antisymmetric Levi-Civita symbol.)
	The constants $k_j$ and the function $m$ may be complex but the coordinates $x^1,x^2,x^3$ and the corresponding momenta are assumed to be real. 
	
	To summarize our problem, we assume that 
	\begin{equation}\label{IC SI}
		[X_1,H]=[X_2,H]=[X_1,X_2]=[Y,H]=0
	\end{equation}
	with the Hamiltonian \eqref{HamMagn}, and integrals \eqref{cyl integrals} and \eqref{int Y}. As mentioned above, we can separate these equations into coefficients of derivatives such as $\pd_{x_1x_3}$ (or $\pd_{rZ}$ in cylindrical coordinates) that must vanish separately. We thus obtain the determining equations which we solve for the fields $\vec{B},W$ together with functions $s^{r,\phi,Z}_{1,2},m_{1,2},m$ and constants $k_j$ in the integrals. (We note that if the constants $k_6$ and $k_3$ are not intertwined with any other constant $k_j$ in \eqref{int Y}, i.e. all other constants $k_j$ in \eqref{int Y} can be set to 0, then the corresponding integrals, $\tilde{X}_1=L_3^A+\tilde{m}(x^1,x^2,x^3)=p_\phi^A+\tilde{m}(r,\phi,Z)$ and $\tilde{X}_2=p_3^A+\tilde{m}(x^1,x^2,x^3)=p_Z^A+\tilde{m}(r,\phi,Z)$, respectively, can be seen as a ``square root'' of the cylindrical integrals \eqref{cyl integrals}. In this case the cylindrical integrals \eqref{cyl integrals} are dependent on these first order integrals and we do not have the required superintegrability.)
	
	We proceed in cylindrical coordinates. We substitute the partially reduced form of magnetic field $\vec{B}$ from \eqref{Bcond} into the equations for $Y$, which we use to constrain $\vec{B}$ further and determine $W$. In the process we split into several cases depending on whether the constants~$k_j$ vanish or not, and subsequently use conditions on $X_1,X_2$ to constrain the fields further and determine their final form. We include the details of our calculation in the Appendix \ref{app}.
	
	Taking into account that the cases with real electromagnetic fields and integrals were considered in O. Kub\r{u}'s Master thesis \cite{Kubu2020}, we are mainly interested in solutions not found there, especially those with complex constants $k_j$ in the integral $Y$.
	
	We find one such system, which we analyze in Subsection \ref{sec new sys}. We also confirm that the systems found earlier can be extended into complex systems by simply allowing complex coupling constant. We analyze those in Subsection \ref{sec old systems}.
	
	\section{Results}\label{sec res}
	\subsection{The new system}\label{sec new sys}
	The corresponding electromagnetic field in Cartesian coordinates is given by 
	\begin{equation}\label{new cyl}
		\vec{B}(\vec{x}) =\left(-\frac{i b}{(x^1- i x^2)^3}, -\frac{b}{(x^1- i x^2)^3}, 0\right),\quad W(\vec{x})=\frac{w_1}{2(x^1- i x^2)^2} - \frac{b^2}{8(x^1- i x^2)^4},
	\end{equation}
	where both coupling constants $b$ and $w_1$ may be complex.
	
	With complex magnetic field, we no longer have the gauge freedom. Let us write all the results in terms of magnetic momenta $p_j^A$.
	
	This system is maximally superintegrable. We have 3 first order integrals,
	\begin{equation}
		Y_1=i p^A_1+p^A_2,\quad Y_2=L^A_2+i L^A_1-\frac{b}{(x^1-i x^2)},\quad \tilde{X}_2=p^A_3+\frac{b}{2(x^1- i x^2)^2},
	\end{equation}
	i.e. one of the cylindrical integrals reduces to a first order one, $X_2=\tilde{X}_2^2$, followed by the Hamiltonian and the second cylindrical integral
	\begin{align}
		H={}&\frac{1}{2}((p^A_1)^2+(p^A_2)^2+(p^A_3)^2)+\frac{w_1}{2(x^1- i x^2)^2} - \frac{b^2}{8(x^1- i x^2)^4},\label{ham}\\
		X_1={}& (L^A_3)^2-\frac{b (x^1+i x^2)}{x^1 -i x^2} p^A_3 -b^2 \frac{(x^1+i x^2)}{2(x^1 -i x^2)^3} +w_1\frac{(x^1+i x^2)}{x^1 -i x^2}.
	\end{align}
	
	The algebraic independence of all 5 integrals is obvious once we notice that $L_3^2$ contains $(x^3)^2 p_1^2$ and $x^3$ is absent from the other integrals.
	
	The determining equations for the classical and quantum (super)integrability are very similar if we choose a suitable form of symmetrization. For first order integrals there is no change at all, for second order integrals only the zeroth order equation differs, namely there is an a priori nonvanishing additional term proportional to $\hbar^2$ \cite{Marchesiello2015}. In \cite{Kubu2020} we showed that the correction for the integral $X_1=L_3^2+\ldots$ has the following form $-\frac{\hbar^2}{2}(x^1\pd_{x^2}-x^2\pd_{x^1}) B^3$ and vanishes for the Cartesian integrals $P_i^2+\ldots$, including $X_1=P_3^2+\ldots$ In our case $B^3=0$, therefore the integrals have the same form in classical mechanics as well. However, interpretation of our system in that context is unclear due to complex coefficients in \eqref{new cyl}.
	
	Comparing the form of the Hamiltonian \eqref{ham} with Case IIc in \cite{Shapovalov1972}, we see that the system separates in cylindrical coordinates, with the corresponding commuting integrals $\tilde{X}_1$ and $X_2$.
	
	Let us introduce ``complex'' coordinates for the $x^1 x^2$ plane, $z=x^1+i x^2$, $\bar{z}=x^1-i x^2$. The corresponding magnetic momenta read
	\begin{equation}
		p^A_z=\frac{1}{2}(p^A_1-i p^A_2),\quad p^A_{\bar{z}}=\frac{1}{2}(p^A_1+i p^A_2).
	\end{equation}
	We note that the vanishing $B^z$, see \eqref{new cyl}, implies that $p_1^A$ and $p_2^A$ commute. The same clearly holds also for $p_z^A$ and $p_{\bar{z}}^A$.
	
	The transformed electromagnetic field reads
	\begin{equation}\label{new cyl com}
		B =\frac{b}{\bar{z}^3} \d \bar{z}\wedge\d x^3,\quad W(z,\bar{z},x^3)=\frac{w_1}{2\bar{z}^2} - \frac{b^2}{8\bar{z}^4},
	\end{equation}
	and the integrals are as follows
	\begin{align}
		Y_1={}&2i p^A_z,\quad Y_2=2 x^3 p^A_{z}-\bar{z} p^A_3- \frac{b}{\bar{z}},\quad \tilde{X}_2=p^A_3+\frac{b}{2\bar{z}^2},\\
		H={}&\frac{1}{2}(4 p^A_z p^A_{\bar{z}}+(p^A_3)^2)+\frac{w_1}{2\bar{z}^2} - \frac{b^2}{8\bar{z}^4},\label{ham comp}\\
		X_1={}& -(z p^A_z-\bar{z}p^A_{\bar{z}})^2-\frac{b z}{\bar{z}} p^A_3-b^2 \frac{z}{2\bar{z}^3} +\frac{w_1 z}{\bar{z}}.
	\end{align}
	The minus sign in $X_1$ is obtained from the complex unit $i$, as the angular momentum $L^A_3=x^1p^A_2-x^2p^A_1$ transforms into $i(z p^A_z-\bar{z}p^A_{\bar{z}})$.
	
	The forms of the Hamiltonian and the integral $X_1$ imply that this system can be seen as an extension of a 2D system (at least in the classical context), a method to obtain 3D system with magnetic field from 2D systems discussed in classical setting without complex coordinates in \cite{Marchesiello2019,Kubu_2021}. There we can set the integral of motion $\tilde{X}_2$ to a constant $\kappa$ and obtain the corresponding 2D system by merging the first and zeroth order terms. The corresponding 2D system lacks the linear term and thus has no magnetic field. We see that the use of complex coordinates $z=x^1+i x^2$, $\bar{z}=x^1-i x^2$ in the 2D system does not interfere with the method.
	
	However, this method is formal in the quantum context because the purely continuous spectrum of $p_3$ precludes a reasonable choice of the constant $\kappa$.
	
	Let us analyze the algebra of integrals. The commutators are as follows, with only the leading order terms for the dependent integrals $R_1$ and $R_2$ written for brevity: 
	\begin{align}
		[Y_1,Y_2]={}&0,\quad [Y_1,\tilde{X}_2]=0, \quad [Y_2,\tilde{X}_2]=-2 i \hbar p_z=\hbar Y_1,\quad 		[X_1,\tilde{X}_2]=0\\
		[Y_1,X_1]={}&2 i\hbar(i \{p_z^A, zp_z^A-\bar{z}p_{\bar{z}}^A\})
		+\ldots=:i\hbar R_1,\\
		[Y_2,X_1]={}& i \hbar \{2 x^3 p_{z}^A-\bar{z} p_3^A ,zp_z^A-\bar{z}p_{\bar{z}}^A\}
		+\ldots=:i \hbar R_2,
	\end{align}
	where $\{\cdot,\cdot\}$ denotes the anticommutator, $\{a,b\}=ab+ba.$
	The higher order commutators are as follows
	\begin{align}
		[Y_1,R_1]={}&- 2 i\hbar Y_1^2,\quad 	[Y_2,R_1]=-i \hbar \{Y_1, Y_2\}, \quad [\tilde{X}_2,R_1]=0,\quad \\
		[Y_1,R_2]={}&-i\hbar\{Y_1, Y_2\},\quad [Y_2,R_2]=-i\hbar(2 Y_2^2-6 b \tilde{X}_2 +2 w_1),\quad [\tilde{X}_2,R_2]=-\hbar R_1,\\
		[X_1,R_1]={}&i \hbar (2\{X_1, Y_1\}-\hbar^2 Y_1),\quad [X_1,R_2]= i \hbar(2 \{Y_2, X_1\}-\hbar^2 Y_2),\label{h3}\\
		[R_1,R_2]={}&i \hbar(4 i b H - 6 i b \tilde{X}_2^2 + 4 i w_1 \tilde{X}_2),
	\end{align}
	so the algebra closes. We notice that the terms with $\hbar^3$ in \eqref{h3} do not arise in the classical mechanics, i.e. they modify the Poisson brackets.
	
	A similar 2D system arises as $E_8$ with $a_2=0$ in the classification of 2D superintegrable systems (see e.g. the review article \cite{Miller2013})
	\begin{equation}\label{e8}
		H=4p_zp_{\bar{z}}+a_1 z\bar{z}-\frac{a_3}{\bar{z}^2},
	\end{equation}
	(notice the choice $m=\frac{1}{2}$ instead of $m=1$ in \cite{Miller2013}),
	was recently analyzed from the point of view of dynamical symmetry algebras \cite{Marquette2022a}. We did not find this more general 2D system, which can be extended into 3D as well, due to our first order additional integral ansatz. In contrast to the system \eqref{e8}, which exhibits an equidistant, oscillator-like spectrum (determined using ladder operators \cite{Marquette2022a}), for our system \eqref{new cyl com} the quadratic integral $p_z^2+\frac{a_1}{4}\bar{z}^2$ reduces to a first order one $p_z=-i\hbar\pd_z$ and there are no ladder operators. The free motion in the $z$ direction of our system suggests a purely essential spectrum. 
	
	\subsubsection{Physical interpretation of the system}
	There are several problems with the rigorous definition of the Hamiltonian \eqref{ham}. First the choice of the vector potential $A$: we no longer have the freedom to choose the imaginary part as the gauge transformation is unbounded. Second, if we cannot choose $A$, we must retain the $(x^1- i x^2)^{-4}$ term in $W$ \eqref{new cyl}. Due to its strong singularity, it cannot serve as a perturbation with respect to magnetic Laplacian $(p^A_1)^2+(p^A_2)^2+(p^A_3)^2$ in the standard definition.
	
	We leave these problems as an open challenge. In this article, we proceed formally by choosing the simplest gauge for \eqref{new cyl} satisfying $\dive \vec{A}=0$, i.e.
	\begin{equation}
		\vec{A}(\vec{x})=\left(0,0,-\frac{b}{2(x^1- i x^2)^2}\right),
	\end{equation}
	which also eliminates the most singular term in $W$ \eqref{new cyl}.
	
	The Hamiltonian \eqref{ham} now reads
	\begin{equation}\label{ham fixed}
		H=\frac{1}{2}(p_1^2+p_2^2+p_3^2) - \frac{b}{2(x^1- i x^2)^2} p_3+ \frac{w_1}{2(x^1- i x^2)^2}.
	\end{equation}
	This Hamiltonian is pseudo-Hermitian \cite{Mostafazadeh2002}, i.e. $\Theta H\Theta^{-1}=H^\dagger$, with $\Theta=\mathcal{P}_2$ a Hermitian (i.e. bounded and self-adjoint) invertible operator that changes the sign of $x^2\to -x^2$ and $p_2\to-p_2$. However, the partial parity $\mathcal{P}_2$ is not a positive operator, therefore our Hamiltonian is not quasi-Hermitian in the sense of \cite{Scholtz1992,dieudonne1961quasi}.
	
	The Hamiltonian is neither invariant under the parity $\mathcal{P}$ nor the time reversal $\mathcal{T}$. It is $\mathcal{PT}$-self-adjoin without being self-adjoint with respect to $\mathcal{P}$ or $\mathcal{T}.$
	
	We therefore conclude that the Hamiltonian \eqref{ham} is pseudo-Hermitian but not quasi-Hermitian and the physical interpretation of this system is not clear.
	
	The Schr\"odinger equation $H\Psi=E\Psi$ for the system \eqref{ham fixed} can be solved in the complex coordinates $z=x^1+i x^2$, $\bar{z}=x^1-i x^2$ by separation of variables with the following smooth non-normalizable solution
	\begin{equation}\label{psi new}
		\Psi \left(z , \bar{z},x^3\right) \sim
		N \exp\left(C z+ \frac{\lambda_3^2-2 E}{4 C \hbar^2}\bar{z}+\frac{w_1- \lambda_3 b}{4 C \hbar^2} \frac{1}{\bar{z}}\right)\exp\left(\frac{i}{\hbar}\lambda_3 x^3\right),
	\end{equation}
	where $C$ is the separation constant and $\lambda_3\in\R$ corresponds to the momentum $p_3$ by the Fourier transform. 
	
	We are not able to prove that the spectrum is purely essential. Both exponentials in \eqref{psi new} need to be purely imaginary  for standard construction of the Weyl sequence, but we do not know whether $\lambda$ is real or not.
	In the oscillating case \eqref{e8} the eigenfunctions $\psi$ are bounded by the Gaussian term $\exp(-\frac{\lambda}{2} z\bar{z})=\exp[-\frac{\lambda}{2} ((x^1)^2+(x^2)^2)]$ so the previously discussed condition is not necessary.
	
	\subsection{Systems found earlier admit complex coupling constants}\label{sec old systems}
	All other systems have already been found in \cite{Marchesiello2015} or \cite{Marchesiello2019} (and considered also in \cite{Kubu2020}), i.e. the form of their integrals is a standard one. We have the following 3 systems, where the coupling constants $b_j,w_j$ may be complex.
	
	\begin{enumerate}
		\item \label{kns W0} The constant magnetic field and vanishing scalar potential
		\begin{equation}
			\vec{B}(\vec{x})=(0,0,b),\quad W(\vec{x})=0,
		\end{equation}
		i.e. our Hamiltonian is 
		\begin{equation}
			H=\frac{1}{2}((p^A_1)^2+(p^A_2)^2+(p^A_3)^2).	
		\end{equation}
		
		Proceeding formally, we present 2 choices of gauge, the first one suitable for separation in Cartesian coordinates, the second one for the cylindrical ones,
		\begin{equation}\label{gauge kons}
			\vec{A}=(0,b x^1,0),\quad \vec{A}=\left(-\tfrac{b}{2}x^2,\tfrac{b}{2}x^1,0\right).
		\end{equation}
		
		{For a general complex $b=\alpha+i \beta$ the Hamiltonian is symmetric under the parity $\mathcal{P}$ but not $\mathcal{P}$-self-adjoint, $\mathcal{P}H\mathcal{P}=H\neq H^\dagger$, and neither $\mathcal{T}$- nor $\mathcal{PT}$-symmetric.}
		
		{	Excluding the self-adjoint real case, the purely imaginary $b=i \beta$ yields an interesting example of $\mathcal{PT}$-symmetric (also $\mathcal{P}$- and $\mathcal{T}$-symmetric) operator which is neither $\mathcal{T}$- nor $\mathcal{P}$-self-adjoint, i.e.
			\begin{equation}
				\mathcal{PT}H \mathcal{PT}=	\mathcal{P}	H \mathcal{P}=\mathcal{T}	H \mathcal{T}=H\neq H^\dagger.
			\end{equation}
			The Hamiltonian is therefore not pseudo-Hermitian with respect to a metric $\mathcal{P}$, $\mathcal{T}$ nor $\mathcal{PT}$, but we cannot exclude a more complicated metric. This is another example showing that $\mathcal{PT}$-symmetry and pseudo- or quasi-Hermiticity are not necessarily equivalent. (In fact, the inequivalence was demonstrated in \cite{Siegl}, see also \cite{KS_book,Krejcirik2015}.)}
		
		This system is maximally superintegrable with 4 first order integrals
		\begin{equation}\label{IP konst pole}
			Y_1=p_1^A-b x^2,\quad Y_2=p_2^A+b x^1,\quad \tilde{X}_1=L_3^A+\frac{b ((x^1)^2+(x^2)^2)}{2},\quad \tilde{X}_2=p_3^A,
		\end{equation}
		and a nonpolynomial one \cite{Marchesiello2015}
		\begin{equation}\label{paty konst pole}
			X_5=(p_2^A+b x^1)\sin\left(\frac{b x^3}{p_3^A}\right)-(p_1^A-b x^2)\cos\left(\frac{b x^3}{p_3^A}\right),
		\end{equation}
		where we assume that $p_3^A\neq0$. Interpretation of the last integral in the quantum mechanical context remains unclear due to the $\frac{x^3}{p_3^A}$ term in the trigonometric functions and purely continuous spectrum of $p_3^A=p_3$.
		
		As in the real case analyzed in \cite{Kubu2020}, the stationary Schr\"odinger equation $\hat{H}\psi=E\psi$ separates in the Cartesian coordinates in the first (Landau) gauge \eqref{gauge kons}
		\begin{gather}\label{red schr}
			\Psi(\vec{x})=f(x)\exp\left(\frac{i}{\hbar}\lambda_2 y\right)\exp\left(\frac{i}{\hbar}\lambda_3 z\right),\\
			\hbar^2 f''(x)=\left((b x-\lambda_2)^2+\lambda_3^2-2E\right)f(x),\\
			Y_2\psi(\vec{x})=\lambda_2\Psi(\vec{x}),\quad \tilde{X}_2\Psi(\vec{x})=\lambda_3\Psi(\vec{x}).
		\end{gather}
		(The separation can be done rigorously using the (unitary) Fourier transform.)
		
		The reduced Schr\"odinger equation for $f(x)$ can be solved in terms of confluent hypergeometric functions \cite{Kubu2020}, which under some condition reduce to Hermite polynomials (see e.g. \cite[Chapter 13]{DLMF}), i.e. the form known from (shifted) 1D harmonic oscillator. This happens if the energy $E$ satisfies
		\begin{equation}\label{E konst}
			E=\frac{\lambda_3^2}{2}+\hbar b\left(n+\frac{1}{2}\right),\quad n=0,1,\ldots,
		\end{equation}
		and the corresponding non-normalizable eigenfunctions are
		\begin{equation}\label{psi n}
			\Psi_{n,\lambda_2,\lambda_3}(\vec{x})=K_{n}\exp\left(\frac{i}{\hbar}(\lambda_2 y+\lambda_3 z)\right) H_n\left(\sqrt{\frac{b}{\hbar}}\left(x-\frac{\lambda_2}{b}\right)\right)\exp\left(-\frac{b}{2\hbar}\left(x-\frac{\lambda_2}{b}\right)^2\right).
		\end{equation}
		The result so far suggests that all the energies are in the essential spectrum, 
		\begin{equation}
			[\hbar b/2,+\infty)\subset\sigma_{ess}(H).	
		\end{equation}
		However, the standard argument, i.e. construction of Weyl sequences, does not work as the last exponential in \eqref{psi n} is not bounded due to the nonvanishing imaginary part ($b=\alpha + i \beta$)
		\begin{equation}
			-\frac{\beta  \left(\alpha^{2} x^{2}+\beta^{2} x^{2}-\lambda_2^{2}\right)}{2 \hbar \left(\alpha^{2}+\beta^{2}\right)}.
		\end{equation}
		We are therefore not able to prove the result above.
		
		\item \label{kns W} Constant magnetic field with a nontrivial scalar potential governing the motion in direction of the $x^3$ axis,
		\begin{equation}
			\vec{B}(\vec{x})=(0,0,b),\quad W(\vec{x})=W_3(x^3),
		\end{equation}
		i.e. our Hamiltonian is 
		\begin{equation}\label{ham konst w}
			H=\frac{1}{2}((p^A_1)^2+(p^A_2)^2+(p^A_3)^2)+W_3(x^3).	
		\end{equation}
		We again continue formally, i.e. we choose the gauge \eqref{gauge kons}. The properties of the Hamiltonian \eqref{ham konst w} with respect the parity $\mathcal{P}$ and/or time-reversal $\mathcal{T}$ carry over from the previous case if and only if $W_3$ has the same property.
		
		In this case the second cylindrical integral remains quadratic,
		\begin{equation}\label{p3 kvadr}
			X_2=\left(p_3^A\right)^2+W_3(x^3),
		\end{equation}
		the other integrals in \eqref{IP konst pole} remain unchanged. The nonpolynomial integral from \eqref{paty konst pole} cannot be used.
		
		Some special choices of the potential $W_3(x^3)$ admit additional integrals, e.g.
		\begin{equation}
			W_3(x^3)=\frac{c}{(x^3)^2}+\frac{b^2 (x^3)^2}{8},\quad W_3(x^3)=\frac{b^2}{2}(x^3)^2,
		\end{equation}
		are known to admit the fifth quadratic integral \cite{Marchesiello2017}. These algebraic results carry over to the case with complex coupling constants, but we again have the problem with rigorous definition of the system.
		
		For \eqref{ham konst w} we also obtain purely essential spectrum consisting of infinitely degenerate energies \eqref{E konst} where we must replace $\lambda_3^2$ by the eigenvalue $\xi$ corresponding to the spectral problem $(\left(p_3^A\right)^2+W_3(x^3)) \psi=\xi\psi$.
		
		\item \label{B y} The system with non-constant magnetic field,
		\begin{align}\label{WB netriv}
			\vec{B}=\left({\frac{4 b}{(x^2)^3}},0,0\right),\quad 	W =-4\left(\frac{b^2}{2 (x^2)^4}	+\frac{w_0}{(x^2)^2}\right),
		\end{align}
		i.e. our Hamiltonian is 
		\begin{equation}
			H=\frac{1}{2}((p^A_1)^2+(p^A_2)^2+(p^A_3)^2)-4\left(\frac{b^2}{2 (x^2)^4}	+\frac{w_0}{(x^2)^2}\right).	
		\end{equation}
		Formally choosing the gauge satisfying $\dive\vec{A}=0$,
		\begin{equation}
			\vec{A}=(0,0,-2b(x^2)^{-2}),
		\end{equation}
		the Hamiltonian is even in $x^2$ and therefore invariant under the parity $\mathcal{P}$. The $\mathcal{PT}$-symmetry then depends on the invariance of constants under time-reversal $\mathcal{T}$, which means that they have to be real. On the other hand, the Hamiltonian is $\mathcal{P}$-pseudo-Hermitian if we choose them purely imaginary. As the parity $\mathcal{P}$ is not positive definite, the Hamiltonian is not quasi-Hermitian with the metric $\mathcal{P},\mathcal{T}$ or $\mathcal{PT}$, though we cannot exclude a more complicated metric.
		
		The corresponding first order integrals of motion are 
		\begin{equation}\label{px p_Z}
			\tilde{X}_2=p_3^A+{\frac{2b}{(x^2)^2}}, \quad Y_1=p_1^A.
		\end{equation}
		The other cylindrical integral $X_1$ is of the second order and reads
		\begin{equation}
			X_1=\left(L_3^A\right)^2 -{\frac{4b \left((x^1)^2 +(x^2)^2 \right)}{(x^2)^2} \left(p_3^A+\frac{2b}{(x^2)^2}\right)}-{\frac{8 w_0\left((x^1)^2 +(x^2)^2 \right)}{(x^2)^2}},\\
		\end{equation}
		This case in the chosen coordinates corresponds to the Case Ib) or Ic) in \cite{Marchesiello2019} with constants
		\begin{equation}
			a_1=a_2=0,\quad a_3=2b,\quad b_1=b_2=0,\quad b_3=-4 w_0.
		\end{equation}
		There is, therefore, another second order integral of motion, namely
		\begin{align}
			X_3&=L_3^A p_2^A
			-\frac{4b x^1}{(x^2)^2}	\left(p_3^A+{\frac{2b}{(x^2)^2}} \right)-	\frac{8 w_0 x^1}{(x^2)^2},
		\end{align}
		but it is not independent due to the relation
		\begin{equation}
			(\tilde{X}_2^2 + Y_1^2 - 2H)X_1 + X_3^2=-4(\tilde{X}_2 b + 2 w_0)(2H-\tilde{X}_2^2).
		\end{equation}
		
		The solution of the stationary Schr\"odinger equation in terms of Bessel functions \cite[equation (2.147)]{Kubu2020} does not yield a dense set of orthogonal polynomials. We have not obtained the full spectrum; we expect a purely essential one but are not able to prove this result.
	\end{enumerate}

	\section{Conclusion}\label{sec: concl}
	Motivated by the recent observation that the complex magnetic field is physically relevant at least in statistical mechanics \cite{Peng2015}, in this article we looked for the first order superintegrable systems of the cylindrical type that admit complex-valued magnetic field, i.e. $B^j:\R^3 \to \C$. 
	
	We have obtained four such systems, three of them known from \cite{Kubu2020} but allowing complex coupling constants and one new system \eqref{new cyl}, where also the integrals of motion feature complex coefficients.
	
	Drawing on results from \cite{Kubu2020}, all these systems separate in cylindrical coordinates and at least one further coordinate system, namely the Cartesian or the extension of ``complex'' coordinate system $z=x^1+i x^2, \bar{z}=x^1-ix^2,x^3$ for the new system. This is a typical behavior for first order superintegrable systems with magnetic field if the integrals have the standard form. On the other hand, the superintegrable system from \cite{Kubu2022} has a nonstandard set of first order integrals and does not separate in any coordinates system.
	
	The new system is maximally superintegrable with first order integrals, the Hamiltonian and one quadratic cylindrical integral. The algebra of integrals closes with two additional quadratic integrals, dependent on the previously mentioned.
	
	The rigorous definition of these systems poses a formidable challenge. In the standard Hermitian setting, the calculations are performed in the gauge-covariant form. With complex magnetic fields we are in the non-Hermitian setting where the gauge freedom no longer holds, and it is not clear how we should choose the corresponding vector potential.
	Leaving the resolution of this problem as an open challenge, we proceeded by choosing the simplest gauge that also eliminates the most singular terms.
	
	Even in this case we face another problem: The spectra of the found systems are not guaranteed to be real (and measurable) as the resulting Hamiltonians are usually only pseudo-Hermitian and therefore self-adjoint only with respect to an indefinite scalar product. Standard methods used to infer the spectrum do not work in this setting. Unlike in the quasi-Hermitian case, where the modified scalar product is positive definite, we thus cannot interpret these systems as nonstandard representations of the usual quantum mechanics.
	
	\section*{Acknowledgements}
	OK thanks David Krej\v{c}i\v{r}k for encouragement to investigate the topic of complex magnetic fields and discussions about the spectral analysis of the found systems.
	
	Computations in this paper were performed using Maple\textsuperscript{TM} 2022.1 by Maplesoft, a division of Waterloo Maple Inc., Waterloo, Ontario.
	
	OK's research is supported by the Grant Agency of the Czech Technical University in Prague, grant No. SGS22/178/OHK4/3T/14.
	
	\appendix
	\section{Detailed solution to the determining equations}\label{app}
	{The condition $[Y,H]=0$ separates into coefficients of the second, first and zeroth order derivatives. The first order ones can be solved for the partial derivatives $\pd_\alpha m,\alpha\in \{r,\phi,Z\}$. Assuming sufficient smoothness, we use the compatibility conditions 
		\begin{equation}\label{comp cond}
			\pd_{\alpha \beta}m =\pd_{\beta \alpha}m,\quad \alpha,\beta\in\{r,\phi,Z\}
		\end{equation}
		to arrive at the division into subcases. Namely, differentiating $\pd_{r\phi}m =\pd_{\phi r}m$ twice with respect to $Z$, we get the simple condition}
	\begin{equation}
		(k_4 \sin (\phi)r-k_5\cos(\phi)r+k_3) r^4 \mu'''(Z)=0.
	\end{equation}
	Together with the condition derived by differentiating $\pd_{rZ}m =\pd_{Z r}m$ with respect to $Z$ twice, namely
	\begin{align}
		\left(\left({k_4} Z -{k_2} \right) \mu''' \left(Z \right)+4 {k_4}\mu'' \left(Z \right) \right) \cos \left(\phi \right)+\left(\left({k_5} Z +{k_1} \right) \mu'''\left(Z \right)+4 k_5\mu'' \left(Z \right) \right) \sin \left(\phi \right)=0,
	\end{align}
	we find that the assumption $k_4\neq0$ leads to $\mu(Z)=\mu_1 Z +\mu_2$. The case $k_5\neq0$ is equivalent to $k_4\neq0$ by a rotation of the coordinate axes.
	
	If both $k_4=k_5=0$, we similarly consider $\pd_Z(\pd_{r\phi}m =\pd_{\phi r}m)$ and $\pd_Z(\pd_{rZ}m =\pd_{Z r}m)$, which read
	\begin{equation}
		-2 k_3 \mu'' \! \left(Z \right) r^{4}=0	, \quad
		(k_1 \sin \! \left(\phi \right) -k_2 \cos \! \left(\phi \right))\mu'' \! \left(Z \right) r^{4}=0.
	\end{equation}
	Eliminating all three constants here would leave only $k_6$, which implies that the integral $Y$ is a reduced version of $X_1$, not an additional independent integral. We can therefore conclude that 
	\begin{equation}
		\mu \left(Z \right) =\mu_1 Z +\mu_2
	\end{equation}
	holds in both cases.
	
	Differentiating $\pd_{r\phi}m =\pd_{\phi r}m$ with respect to $r$ and $Z$ now yields
	\begin{equation}
		(\rho'''(r)r^2+2\rho''(r)r -2 \rho(r))(k_4\sin(\phi)-k_5\cos(\phi))=0,
	\end{equation}
	We therefore split our considerations into two cases: 
	\begin{enumerate}
		\item \label{k40} $k_4=k_5=0$, 
		\item \label{k4neq0} $k_4\neq0$ without loss of generality, i.e. $\rho(r)=\rho_3 r^2 +\rho_1+\rho_2 r^{-1}$.
	\end{enumerate}
	
	\subsection{Case \ref{k40} - $k_4=k_5=0$}
	In this case we shall consider two complementary possibilities: $k_1\neq 0$ or $k_2 \neq 0$, and $k_1=k_2=0$ with $k_3=c k_6$ a nontrivial combination, i.e. $c\neq0$ and $k_6\neq0$.
	
	Let us treat the latter case first. It does not yield anything new in the end. Substituting the assumptions $k_1=k_2=0$ and $k_3=c k_6$ into the compatibility conditions \eqref{comp cond}, we immediately see that the magnetic field $B$ depends only on $r$. A straightforward calculation using the equations for $Y$ and the first order equations from $[X_1,H]=0, [X_2,H]=0$ and $[X_1,X_2]=0$ yield that also the scalar potential $W$ depends on $r$ only, which means the coordinates $\phi$ and $Z$ are cyclic, e.g. we get no additional integral, only the reduced cylindrical ones needed for integrability.
	
	We can therefore proceed with $k_1\neq0$ without loss of generality. (This is equivalent to $k_2\neq0$ by a rotation of the coordinate system.) The compatibility conditions \eqref{comp cond} differentiated with respect to $r$ twice yield
	\begin{equation}
		\rho \left(r \right) = \frac{ {\rho_1} r + {\rho_2} \ln \left(r \right) r + {\rho_3} r^{3}+ {\rho_4}}{r},
		\quad \sigma \left(r \right) = \frac{ {k_2} \mu_1 r^{2}}{2 {k_1}}+\frac{ {\sigma_1}}{2 r}+\frac{ {\sigma_2}}{6 r^{2}}+ {\sigma_3} r + {\sigma_4}.
	\end{equation}
	Plugging this result back into the compatibility conditions, we immediately find that $\sigma_3=0$ and $k_3 \mu_1=0$.
	
	When we choose $\mu_1\neq0$, the compatibility conditions for $m$ can be solved for the remaining auxiliary functions $\tau(\phi)$ and $\psi(\phi)$, but the remaining equations for integrals $X_1$ and $X_2$ lead to a contradiction.
	
	We therefore continue with $\mu_1=0.$ The terms without $r$ in the compatibility conditions for $m$ can be solved for $\psi(\phi)$ and $\tau(\phi)$ in general as they do not contain $k_3$ nor $k_6$. The last remaining terms in the same equations yield further constraints on the constants in these functions, leading to 2 possible magnetic fields (absorbing $\rho_3$ into $\mu_2$ by a redefinition),
	\begin{equation}\label{nonconst B}
		\begin{split}
			{B^Z} \left(r , \phi , Z\right) ={}& 
			\mu_2r -\frac{2 k_1^{2} {\psi_2} +2 k_1 k_2 {\psi_4} +2 k_2^{2} {\psi_3}- k_2^{3} \rho_4}{2 \left( k_2 \cos \left(\phi \right)- k_1 \sin \left(\phi \right)\right)^{3}r^2},\\
			{B^\phi} \left(r , \phi , Z\right) ={}& 
			-\frac{\left( k_1^{2}+ k_2^{2}\right) {\sigma_2} -12 {\tau_1}}{12 \left( k_2 \cos \left(\phi \right)- k_1 \sin \left(\phi \right)\right)^{2} r^{3}},\\
			{B^r} \left(r , \phi , Z\right) ={}& 
			-\frac{\left(\left( k_1^{2}+ k_2^{2}\right) {\sigma_2} -12 {\tau_1} \right) \left( k_1 \cos \left(\phi \right)+ k_2 \sin \left(\phi \right)\right)}{12 \left( k_2 \cos \left(\phi \right)- k_1 \sin \left(\phi \right)\right)^{3} r^{2}},
		\end{split}
	\end{equation}
	where $\tau_1,\psi_i$ are constants in $\tau(\phi)$ and $\psi(\phi),$
	and 
	\begin{align}\label{const B}
		B^Z(r, \phi, Z) ={}& \mu_2 r,\quad B^\phi(r, \phi, Z) = 0,\quad B^r(r,\phi, Z) = 0.
	\end{align}
	Let us treat these cases separately.
	
	\subsubsection{Nonconstant magnetic field}
	As we have noted, the compatibility conditions \eqref{comp cond} have been solved, so we can obtain the form of $m$ from the first order equations for integral $Y$. The first order equations for $X_2$ now yield the function $m_2$, and imply $\mu_2=0$ with $W=W(r,\phi)$, which the zeroth order equation for $Y$ reduces to a function of one variable,
	\begin{equation}\label{W xy}
		W=W(k_1 r \sin(\phi)-k_2 r \cos(\phi))=W(k_1y-k_2 x).
	\end{equation}
	
	At this moment the equations for $Y$ and $X_2$ are solved. The commutation $[X_1,X_2]=0$ is solved once we set $m_1(r,\phi,Z)=m_1(r,\phi)$, so only the equation $[X_1,H]=0$ remains to be solved.
	
	The coefficient of $\pd_Z$ in $[X_1,H]=0$ is now only a constraint on the constants. We obtain 2 solutions, both yield a magnetic field rational in $y$ after transforming into Cartesian coordinates. The coefficient of $\pd_r$ in $[X_1,H]=0$ reduces $m_1(r,\phi)$ further. The last first order equation, coefficient of $\pd_\phi$, and the zeroth order equation both contain $W$. For them to be compatible with \eqref{W xy}, the magnetic field reduces to the form considered in Case \ref{B y} of Subsection \ref{sec old systems} or to a constant magnetic field, which we treat in the following subsection.
	
	\subsubsection{Constant magnetic field}	
	Here the calculations proceed in the same way as for real magnetic field, see e.g. \cite{Kubu2020}. There is no problem to solve all first order equations, which yield the functions $m,m_1,m_2$ as well as the separation of $W$, 
	\begin{equation}
		W(r,\phi,Z)=W_1(r)+\frac{W_2(\phi)}{r^2}+W_3(Z).
	\end{equation}
	The zeroth order equation for $Y$ now reads
	\begin{multline}\label{70}
		\left({k_1} \cos \! \left(\phi \right)+{k_2} \sin \! \left(\phi \right)\right)W_1'(r)+k_3 W_3' \! \left(Z \right)	+\frac{{k_6} W_2' \! \left(\phi \right)}{r^{2}}\\
		-\frac{\left({k_1} \sin \left(\phi \right)-{k_2} \cos \left(\phi \right)\right) W_2' \left(\phi \right)+2 \left({k_1} \cos \! \left(\phi \right)+{k_2} \sin \! \left(\phi \right)\right) {W_2} \! \left(\phi \right)}{r^{3}}=0.
	\end{multline}
	Zeroth order equation for $X_1$ reads
	\begin{multline}\label{eq}
		W_2'(\phi)\mu_2	+	\left({c_1} \cos \! \left(\phi \right)+{c_2} \sin \! \left(\phi \right)\right)W_1'(r)+c_3 W_3' \! \left(Z \right)	+\frac{{c_6} W_2' \! \left(\phi \right)}{r^{2}}\\
		-\frac{\left({c_1} \sin \left(\phi \right)-{c_2} \cos \left(\phi \right)\right) W_2' \left(\phi \right)+2 \left({c_1} \cos \! \left(\phi \right)+{c_2} \sin \! \left(\phi \right)\right) {W_2} \! \left(\phi \right)}{r^{3}}=0,
	\end{multline}
	where $c_i$ are redefined constants from the auxiliary functions $\rho(r),\sigma(r),\tau(\phi),\psi(\phi)$ not appearing in the magnetic field, i.e. they appear in the first order terms of integrals $X_1,X_2.$ The assumption $k_1\neq0$ allows us to solve for $W_1(r)$ by multiplying equation \eqref{70} by $r^3$ and differentiating 3 times with respect to $r$. Separating the equations \eqref{70} and \eqref{eq} into coefficients of $r$ now eliminates some constants from $W_1(r)$ and with $\mu_2\neq0$ (nonvanishing magnetic field) yields $W_2(\phi)$ equal to a constant that exactly cancels the $r^{-2}$ term in $W_1(r)$. We conclude that 
	\begin{equation}
		W(r,\phi,Z)=w_1	\text{ or } W(r,\phi,Z)=W_3(Z)
	\end{equation}
	depending on the values of constants $k_3$ and $c_3$. The constant in $w_1$ in the first case is irrelevant, so we set it to zero.
	
	We therefore arrived at systems analyzed in \cite{Kubu2020,Marchesiello2019} and listed as Cases \ref{kns W0} and \ref{kns W}, respectively, in Subsection \ref{sec old systems}.\medskip
	
	We conclude that the case $k_4=k_5=0$ yields only systems that were already known from \cite{Kubu2020} but with complex coupling constants. We analyzed them further in Subsection \ref{sec old systems}.
	
	\subsection{Case \ref{k4neq0} - $k_4\neq0$}
	Differentiating all compatibility conditions \eqref{comp cond} with respect to $r$ and $Z$, the functions $\rho(r)$ and $\sigma(r)$ are determined to read
	\begin{equation}
		\rho(r)=\rho_3 r^2 +\rho_1+\rho_2 r^{-1},\quad \sigma(r)=\sigma_3+\sigma_1 r^{-2},
	\end{equation}
	and we also find that $\mu_1=0$. 
	
	We plug these results into the compatibility conditions \eqref{comp cond} again. The highest order terms in $r$ imply $\mu_2=\rho_3$. Next, the terms with $Z$ can be solved for $\tau(\phi)$ and $\psi(\phi)$. The remaining terms in the compatibility conditions now contain constants, some of which are therefore eliminated, leading to only one solution with nontrivial magnetic field. Namely, our auxiliary functions read
	\begin{gather}
		\mu\left(Z \right) = \rho_3,\quad \rho(r)=\rho_3 r^2 +\rho_1+\rho_2 r^{-1},\quad \sigma(r)=\sigma_3+\sigma_1 r^{-2},\\
		\psi\left(\phi \right) = 2\psi_3\cos(\phi) +i\psi_2 e^{i\phi}+\rho_2,\quad \tau(\phi)=\sigma_1-\tau_1 e^{2\phi},
	\end{gather}
	and the corresponding magnetic field reads
	\begin{equation}\label{mag}
		B^r=-i \frac{\tau_2 e^{2i\phi}}{r^2},\quad B^\phi = -\frac{\tau_2 e^{2i\phi}}{r^3},\quad B^Z= 0.
	\end{equation}
	We thus assume $\tau_2\neq0$ in this subsection.
	
	The compatibility conditions for $m$ are now solved. Substituting the auxiliary functions into the first order equation for $Y$ yields
	\begin{equation}
		m \left(r , \phi , Z\right) = 
		\frac{\tau_2(2i k_4 r e^{i \phi}  + i k_3 e^{2i\phi})}{2 r^{2}}.
	\end{equation}
	
	We proceed to solve the first order equations for $X_1$. Those corresponding to $\pd_\phi$ and $\pd_r$ yield $m_1$ and $W$. We subsequently substitute the results into the equation corresponding to $\pd_Z$, differentiate it with respect to $r$ and obtain the terms multiplying $e^{2i\phi},e^{3i\phi}$
	\begin{equation}
		\frac{8 \left(i \psi_2 +\psi_3 \right) \tau_2}{r},\quad 2 i \rho_1 \tau_2,
	\end{equation} 
	which due to our assumption $\tau_2\neq0$ imply $\rho_1=0$ and $\psi_2=i \psi_3$.
	
	The zeroth order equation for $X_1$ and $Y$ yield the final form of the scalar potential
	\begin{equation}
		W(r,\phi,Z)=\frac{w_1 e^{2 i \phi}}{2 r^2}-\frac{\tau_2^{2} e^{4i \phi}}{8 r^{4}}.
	\end{equation}
	The equations for $X_2$ can be solved for $m_2$ and imply $\rho_3=0$.
	
	Thus we obtain the new system written in Cartesian coordinates \eqref{new cyl} and analysed in Subsection \eqref{sec new sys}, with $\tau_2=b$.
	
	\bibliography{AIP}
	
\end{document}